\documentclass[manuscript]{aastex}

\begin{document}

\title{Is the magnetic field in the heliosheath 
laminar or a turbulent bath of bubbles?}

\author{M. Opher\altaffilmark{1}}
\affil{Astronomy Department, Boston University, 725 Commonwealth
  Avenue, Boston, MA}
\email{mopher@bu.edu}

\author{J. F. Drake\altaffilmark{2}}
\affil{University of Maryland, College Park, MD} 
 
\author{M. Swisdak\altaffilmark{3}}
\affil{University of Maryland, College Park, MD}

\author{K. M. Schoeffler \altaffilmark{3}}
\affil{University of Maryland, College Park, MD}  

\author{J. D. Richardson\altaffilmark{4}}
\affil{Kavli Institute for Astrophysics and Space Research, 
Massachusetts Institute of Technology, Cambridge, MA}

\author{R. B. Decker\altaffilmark{5}}
\affil{John Hopkins Univ., Applied Physics Lab., Laurel MD}

\author{G. Toth\altaffilmark{6}}
\affil{University of Michigan, Ann Arbor, MI}

\begin{abstract}
All the current global models of the heliosphere are based on the
assumption that the magnetic field in the heliosheath, in the region 
close to the heliopause is laminar. We argue that in that
region the heliospheric magnetic field is not laminar but instead
consists of magnetic bubbles.  We refer to it as the bubble-dominated
  heliosheath region. Recently, we proposed 
that the annihilation of the ``sectored'' magnetic field within the
heliosheath as it is compressed on its approach to the heliopause
produces the anomalous cosmic rays and also energetic
electrons. As a product of the annihilation of the sectored magnetic
field, densely-packed magnetic islands (that further interacts to
  form magnetic bubbles) are produced. These magnetic
islands/bubbles will be convected with the ambient flows as the sector
region is carried to higher latitudes filling the 
heliosheath. We further argue that the magnetic
islands/bubbles will develop
upstream within the heliosheath. As a result, the magnetic
field in the heliosheath sector region will be disordered well
upstream of the heliopause. We present a 3D MHD simulation with
very high numerical resolution that captures the north-south
boundaries of the sector region. We show that due to the high pressure 
of the interstellar magnetic field a north-south asymmetry develops 
such that the disordered sectored region fills a large portion of the northern
part of the heliosphere with a smaller extension in the southern
hemisphere.  We suggest that this scenario is supported by the
following changes that occur around 2008 and from 2009.16 onward: a) the sudden decrease in the intensity of low
energy electrons (0.02-1.5MeV) detected by Voyager 2
; b) a sharp reduction in the intensity of fluctuations of the radial flow; and c)
the dramatic differences in intensity trends between
galactic cosmic ray electrons (3.8-59 MeV) at Voyager 1 and 2. We argue that these observations are a consequence of Voyager 2
leaving the sector region of disordered field during these periods and crossing
into a region of unipolar laminar field. 

\end{abstract}


\section{Introduction}

The current understanding of the outer heliosphere has changed
dramatically in the last few years due to the recent observations of Voyager
1 and 2 (Stone et al. 2005, 2008) and IBEX (McComas et al. 2009). The
Voyager data at the crossing of the termination shock (TS) revealed that
the anomalous cosmic rays (ACRs) 
were not accelerated at the location of the spacecraft. Several scenarios
were suggested to explain these surprising observations: the ACRs
could be accelerated by the TS along the flanks of the heliosphere (McComas \& Schwadron
2006), in the heliosheath by compressional
turbulence (Fisk \& Gloeckler 2006, 2007, 2009) or by reconnection in
the sectored field near the heliopause (Lazarian \& Opher 2009, Drake et al. 2010). Another shift
in paradigm occurred when it became clear
that the pickup ions carry a large portion of the plasma energy (Zank et al. 1999, Richardson
et al. 2008). Finally the data from IBEX revealed a ribbon of
energetic neutral atoms (ENAs), suggesting that the
interstellar magnetic field has a very important role in shaping the
heliosphere. This result reinforces the earlier Voyager observations: 
the interstellar magnetic field produces a north-south asymmetry in 
the location of the termination shock 
(Opher et al. 2006, 2009).

Recently, we suggested a new mechanism for the acceleration of the
ACRs. We suggested that the sectored
heliospheric magnetic field, which results from the flapping of the
heliospheric current sheet, piles up as it approaches the heliopause,
narrowing the current sheets that separate the sectors, and triggering
the onset of collisionless magnetic reconnection (Drake et al. 2010). 
We argued that reconnection is responsible for the acceleration of the
ACRs and also energetic electrons. 

We showed (Drake et al. 2010), using particle-in-cell (PIC) simulations, that the
sectors break up into a bath of magnetic islands and that most of the
magnetic energy goes into energetic ions, with significant but smaller
amounts of energy going into electrons. The most energetic ions gain
energy as they reflect from the ends of contracting magnetic islands,
a first-order Fermi process. The simulations also revealed that the 
mirror and
firehose conditions play an essential role in the reconnection
dynamics and particle acceleration. An analytic model was constructed
in which the Fermi drive, modulated by the approach to firehose
marginality, is balanced by convective loss. The ACR differential
energy spectrum takes the form of a power law with a spectral index
slightly above $1.5$.

Here we analyze the global topology of the heliospheric current sheet
(HCS) and the sector region.
We argue that within the sector region but upstream of the heliopause, the
heliospheric magnetic field 
is not laminar
but instead filled with nested magnetic islands. The magnetic
islands/bubbles formed during reconnection of the sector region
upstream of the
heliopause will
be convected with the flows as 
the sector boundary is carried to higher latitudes, filling the heliosheath upstream
of the heliopause. 

We argue that due to the increased pressure of the interstellar
magnetic field (Opher et al. 2006, 2007) the sector region and
embedded islands are carried mostly to 
the northern hemisphere. We therefore predict an asymmetry of the magnetic
structure between the northern and southern hemispheres and between
the heliosheath sectored region and the field outside of it. Therefore we
predict that the northern hemisphere will be predominantly a disordered field, filled with magnetic
islands and not a laminar field. 

We further argue that the magnetic islands might develop upstream (but
still within the heliosheath) where collisionless reconnection is
unfavorable -- large perturbations of the sector structure near the
heliopause might cause compressions of the current sheet upstream,
triggering reconnection. As a result, the magnetic field in the
heliosheath sector region will be disordered well upstream of the
heliopause. If this hypothesis is correct, the Voyagers may already
have crossed into a region of disordered field consisting of nested
magnetic islands. We present data from reconnection simulations of a
sectored magnetic field, using a particle-in-cell code that,
surprisingly, exhibit characteristics similar to the Voyager data: the
magnetic field exhibits reversals but with a more erratic spacing
than the initial state; and reconnection of the nested islands is
suppressed due to the approach to the firehose marginal stability
condition so plasma flows are irregular and only occasionally exhibit
traditional reconnection signatures. We denote the late-time non-reconnecting
magnetic islands as ``bubbles'' since in cross section they more
closely resemble a nested volume of soap bubbles than a system of reconnecting islands.

Energetic electrons are especially sensitive to 
to the large-scale magnetic structure of the heliosheath because of their high velocity. 
The presence of magnetic islands/bubbles in the heliosheath sector
region will change the electron transport. 
Specifically a disordered heliospheric magnetic field
near the heliopause acts as the window through which galactic cosmic
ray electrons travelling along the interstellar magnetic fields can enter and percolate
through the heliosphere. Magnetic islands/bubbles will act as local traps for
these energetic electrons. Since lower energy electrons are produced
during reconnection in the sectored field in the same region, these
two classes of electrons should display similar modulation
characteristics. Therefore, we argue that there should be a north-south asymmetry
between the electron intensities at Voyager 1 and 2 and
between the intensities measured while the spacecrafts are within the 
bubble dominated region and outside of it. 
 
This prediction will change our view of the heliosphere since all the
current global models 
(e.g., Opher et al. 2006, 2009, Heerikhuisen et al. 2009, Pogorelov et
al. 2009, Ratkiewicz et al. 2009) are based on the presumption that the magnetic field in the 
outer heliosheath close to the heliopause (HP) and in the heliosheath
sector region is laminar. More recently Czechwoski et al. 2010
analyzed the behaviour of the HCS in a kinematic model. 
They suggested that close to the HP mixed polarities could reconnect making the field random.

We present Voyager 1 and 2 data that supports this disordered field
hypothesis. The Voyager 1 and 2 are moving through the sectored
region, 
are near the boundary between the sectored and non-sectored region,
and have therefore sampled both regions.

The structure of this article is the following: in the next section we
show the global behavior of a modeled sector region based on a MHD
description. This will set the context for the Voyager trajectories
compared with the sector region. We then present Voyager 1 and 2 observations that support our
  proposed scenario. We then present the results of PIC simulations of the
sector region and discuss the structure of the resulting nested magnetic
islands/bubbles and plasma flows. Finally, we discuss the implications of our
predictions for the understanding of the current measurements by
Voyager 2, and our global understanding of the heliosheath, the sector region and heliopause.

\section{Global Behaviour of the Sector Boundary}

Because of the tilt between the solar rotation and magnetic axis,
the actual HCS oscillates up and down, producing a magnetic field with
a ``ballerina skirt'' shape. The drop of the solar wind speed beyond the termination
shock causes the sectors to compress. This behavior was demonstrated in MHD simulations in Drake et al. 2010 and
with greater resolution by Borovikov et al. 2008. At the current stage
in the solar cycle, the limits of the sector boundary are
approximately $\pm 30^{\circ}$, 
close to the latitudes of Voyager 1 and 2. 

We have carried out simulations with a sector region with a
latitudinal width of $60^{\circ}$. We used a 3D MHD
multifluid model described in Opher et al. 2009. This model is based
on BATS-R-US (for a more recent description see Toth et al. 2011). Our model has five
fluids (similar to Alexashov \& Izmodenov (2005) and Zank et
al.(1996)). In this approach there are four populations of neutral H
atoms, for every region in the interaction between the solar wind and
the interstellar wind. Population 4 represents the H atoms of
interstellar origin. 
Population 1 represents the H atoms that exist in
the region between the bow shock and heliopause. Populations 3 and 2
represent the H atoms in the supersonic solar wind and in the
compressed region between the termination shock and the heliopause,
respectively. All four H populations are described by separate systems
of the Euler equations with the corresponding source terms. Each
  population is created in its respective region but is free to move 
between the different regions as the simulation
  evolves. The
ionized component interacts with the H neutrals via charge
exchange. For more details see Opher et al. (2009). The
parameters for the density, velocity and temperature for the ions and
neutrals in the interstellar medium reflect the best observational
values. The parameters for the inner boundary (located at 30 AU) were chosen
to match those used by Izmodenov et al. (2008): proton density of $n =
8.74 \times 10^{-3} cm^{-3}$, temperature $T = 1.087 \times 10^{5}\,
K$, speed of $v=417 km/s$ and a Parker spiral magnetic field
with strength $B = 7.17 \times 10^{-3} G$ at the equator. The outer boundary conditions
are $n = 0.06 cm^{-3}$, velocity equal to $26.3 km/s$, and
$T = 6519K$. The neutral hydrogen in the local interstellar medium is assumed to have 
$n =0.18 cm^{-3}$ and the same velocity and temperature as the ionized
local interstellar medium. 

The interstellar magnetic field $B_{ISM}$  intensity taken was $4.4\mu G$ and the
orientation was such that the angle between
the interstellar velocity and magnetic field is $20^{\circ}$ and the
angle between the plane containing the interstellar velocity and
magnetic field and the solar equator is $60^{\circ}$.  
This orientation and intensity produced the asymmetries in the termination shock locations as
measured by the Voyagers (Opher et
al. 2009). Matching the heliosheath flows measured by Voyager 2
  requires slightly different values (Opher et al. 2009). This
    chosen orientation is close to the one inferred from IBEX (McComas
    et al. 2010) and the ones currently used based on hydrogen deflection
    arguments (Pogorelov et al. 2008). 

We used fixed inner boundary conditions for the ion and neutral fluids. 
The outer boundaries were all
outflows with the exception of the -x boundary, where the inflow
conditions were imposed for the ionized and the population of neutrals
coming from the interstellar medium (Population 4). 
The outer boundaries of the grid are set at  $-1000 AU$ and $1000 AU$ in x, y, z directions, respectively.
The description of the coordinate system is given in Opher et al. (2009).

To capture the sector boundary in the heliosheath adaptive mesh refinement was
used. Several refinements were done throughout the computational
run. The
final computational cells ranges from $0.03~AU$ to $31.25~AU$.
The block used had 8x8x8 cells (see Toth et al. 2011). 
The total number of blocks was $3.0 \times 10^{6}$ blocks and the
total number of cells was $1.4\times 10^{9}$. The
simulation presented here was run for $83.2\,$ years of simulation time on approximately
2000 CPUs. More details of this simulation and the steps involved are given
in Opher et al. 2011. 

Figure 1 shows the distribution of the magnitude of the magnetic field
in the meridional plane (x-z plane). The sector spacing
decreases downstream of the termination shock and further decreases 
on the approach to the heliopause. The approximate trajectories of Voyager
1 and 2 are indicated with red and blue lines, respectively. Figure 2
shows a blowup of the same meridional cut.

There is an asymmetry of the heliosphere due to the
interstellar magnetic field. The inclination of the interstellar
magnetic field compresses the southern hemisphere, pushing the
heliopause and the termination shock in this region closer to the Sun 
(Opher et al. 2006, 2007, 2009). The effect of the difference in pressure
between the north and south is to deflect the sector boundary to the north. Figures 1 and 2 show this effect.

We are able to resolve the alternating sectors out to the middle of the 
heliosheath. The sector structure is compressed downstream of the TS (the sector spacing decreases from
4.7AU to 1.9AU) and the spacing of the sectors continuously decreases further into the heliosheath
as the radial plasma velocity decreases.
Deep into the heliosheath resolutions finer than $0.03~AU$ are
needed. In any case, the sectors close to the heliopause are expected to reconnect and
possibly trigger reconnection further upstream. The decrease in
spacing of the sectors is seen in Figures 2 and 3. Figure 3 presents a zoom of the sectors between the region just
  upstream of the termination shock into the 
  heliosheath. The decrease of the spacing of the sectors can be seen as the solar wind
  crosses the termination shock. The loss of resolution
  deep in the heliosheath can be seen as well.

In the heliosheath the heliospheric magnetic field, to lowest order,
simply adds to the dominant thermal pressure so the sector structure
is not expected to alter the large-scale plasma flow pattern. Thus, we
can simply follow the flow streamlines to determine what fraction of the
sector  region will be carried to the
northern hemisphere even though the resolution is insufficient to
define the sectors. This same procedure was recently followed by
Czechowski et al. 2010. The white streamlines in Figure 1 show the
 boundary of the sector
region. With increased resolution and a longer simulation time we expect the sector region to
fill the domain bounded by the white streamlines. 
We can see that the northern sector region is much
thicker than the sector region carried to
the south. The thickening of the northern sectored region results from the
increased pressure of the interstellar magnetic field in the south,
which diverts the overall heliosheath flow to the north. Thus, it is
evident from Fig. 1 that a simple straight-line extrapolation of the latitudinal
extent of the sectored region fails deep into the heliosheath, 
especially in the north. This could explain why early in 2009
  Voyager 1 seemed to be in a unipolar region while later in 2009 Voyager
  1 reentered the sector region (Burlaga \& Ness 2010).

\begin{figure}[htbp]
\centering
\includegraphics[angle=90, width=0.9\textwidth]{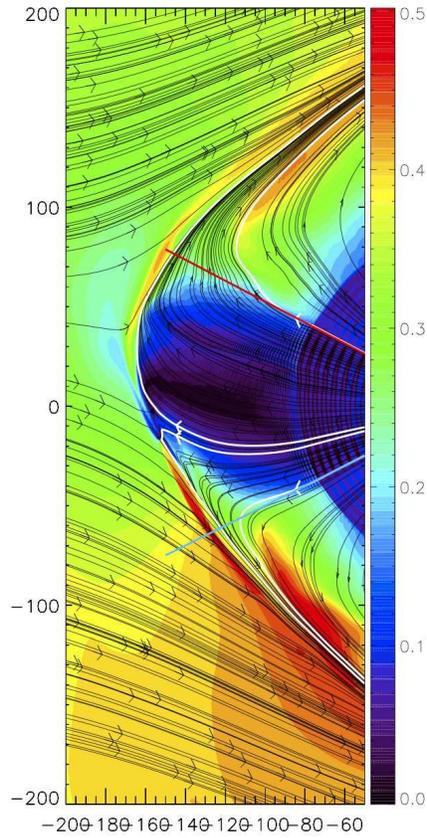}             
\caption{Meridional cut from a 3D MHD simulation showing 
the magnitude of the magnetic field (nT). The sector region of width
of $60^{\circ}$ is the blue-black region. The flow streamlines are
shown in 
black. The boundary of the sector region is shown in the white
streamlines. The Voyager 1 trajectory is $30^{\circ}$ above the solar
equator and that of Voyager 2 is $29.8^{\circ}$ below the solar
equator and are 
shown, respectively, in
the red and blue lines. }
\label{figure1}
\end{figure}

\begin{figure}[htbp]
\centering
\includegraphics[angle=90,width=0.9\textwidth]{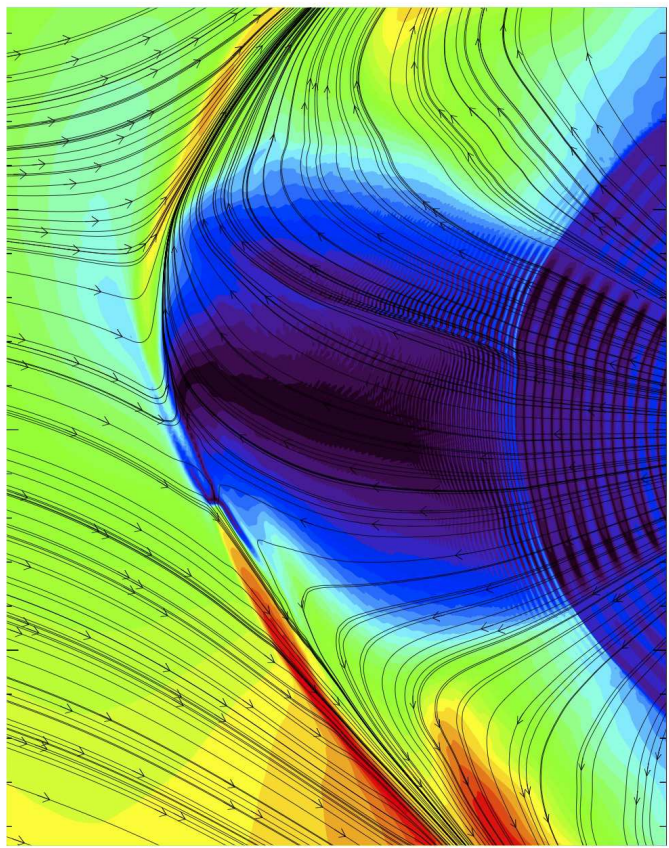}              
\caption{A blowup of the sector region of Figure 1}
\label{figure2}
\end{figure}

\begin{figure}[htbp]
\centering
\includegraphics[angle=90, width=0.9\textwidth]{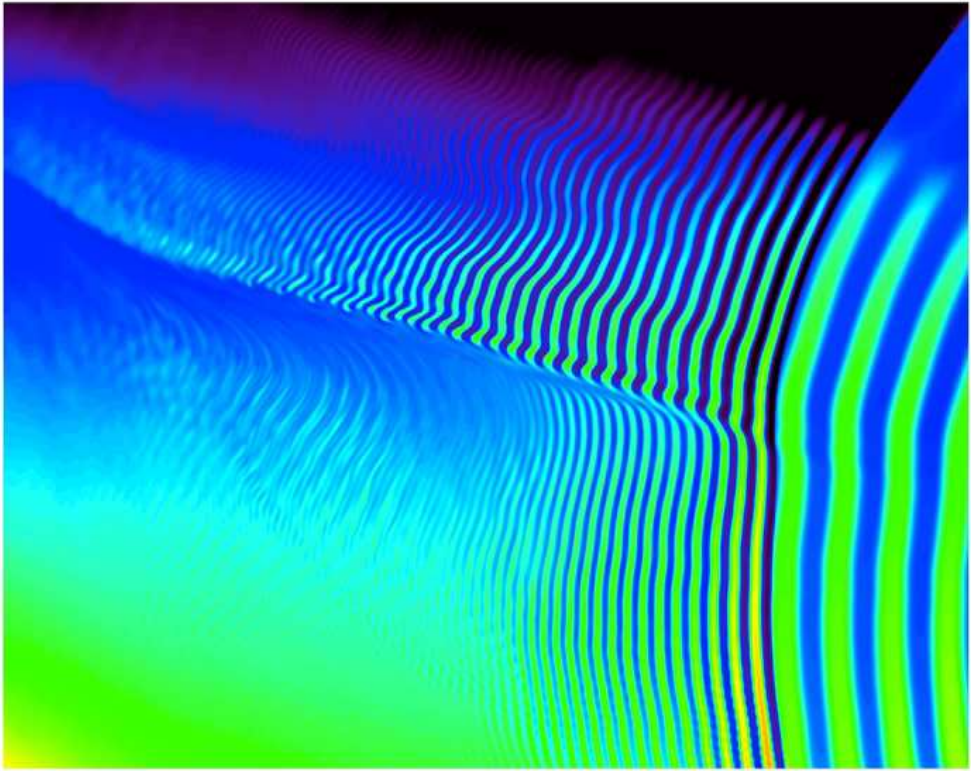}             
\caption{A zoom of the sectors between the region just 
upstream of the termination shock into the 
  heliosheath. The decrease in the spacing of the sectors can be seen 
as the solar wind crosses the termination shock. The loss of resolution
  deep in the heliosheath can be seen as well.}
\label{figure3}
\end{figure}

As mentioned earlier, we predict (Drake et al. 2010) that in the 
heliosheath, near the heliopause the magnetic fields in the sector
region will reconnect, forming nested magnetic islands. We expect the
reconnection to be most robust close to the heliopause. On the other
hand, the perturbations of the sector structure  due to reconnection 
near the heliopause
will cause compressions of the current sheets upstream, possibly triggering
reconnection there. The
heliosheath flows will carry the magnetic islands/bubbles within the sector
region to higher latitudes. Because the dominant
heliosheath flow is northward, we therefore predict an asymmetry of
the magnetic structure between the northern and southern hemispheres. 

\section{Observations of Voyager 1 and 2 that Support our Scenario}

We present Voyager 1 and 2 data that support our disordered
  field scenario in Figures 4, 5 and 6. The shaded gray regions
  are the unipolar regions and the unshaded regions (beyond 2008)
  corresponds to the sector region (the proposed bubble region). In Figure 4 are daily averages of the
intensity of the magnetic field. In the periods of 2008-2008.2 and 2009.15 on (shaded areas) 
Voyager 2 had a positive polarity ($\lambda= 270^{\circ}$) with only
infrequent excursions to negative polarities. In
the time period of 2009.15 on, the field was $90\%$ in the positive
polarity with $10\%$ in the negative polarity. Figure 5 is the daily average of the intensity of electrons from
0.022-0.035 MeV, from 0.035-0.061MeV and 0.35-1.5 MeV. Around the period of 2008-2008.2 and from
2009.15 on (shaded areas), there was a drop of the intensity
of electrons as measured by Voyager 2. The drop was especially
  dramatic after 2009.15. In Figure 6 are 0.2 year
averages and standard deviations of the radial flows, The measured unipolar
regions (between 2008-2008.2 and 2009.15 and 2009.41) are shaded gray
and the conjectured unipolar region (after 2009.41) is shaded light gray. 
We can see that there is a large increase in the intensity of
fluctuations in the sectored region (non-shaded) as compared with the
unipolar regions (shaded). These observations will be discussed more
fully after we present the results of the PIC simulations and associated
discussion. 

\begin{figure}[htbp]
\centering
\includegraphics[angle=90, width=0.9\textwidth]{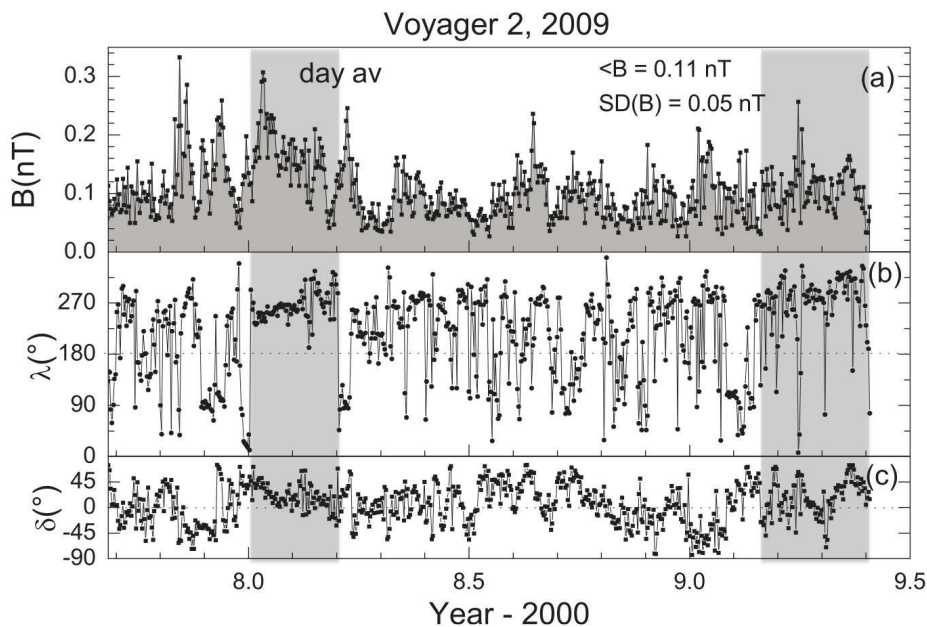}             
\caption{Daily averages of the (a) magnetic strength B, (b) azimuthal
  angle $\lambda$, and (c) elevation angle $\delta$ measured by
  Voyager 2. The azimuthal angle
    $\lambda$ and the elevation angle $\delta$ of B are defined as
    $\lambda=tan^{-1}(B_{Y}/B_{X})$ and $\delta=sin^{-1}(B_{z}/B)$,
    where $B_{X}$, $B_{Y}$, and $B_{Z}$ are the magnetic field components in
    the heliographic coordinate system. The unipolar periods between
    2008-2008.2 and 2009.15 on, are shaded in gray. The non-shaded regions
      correspond to the sector region (proposed ``bubble''
      region.) (Courtesy of L. Burlaga.) }
\label{figure4}
\end{figure}

\begin{figure}[htbp]
\centering
\includegraphics[angle=270, width=0.9\textwidth]{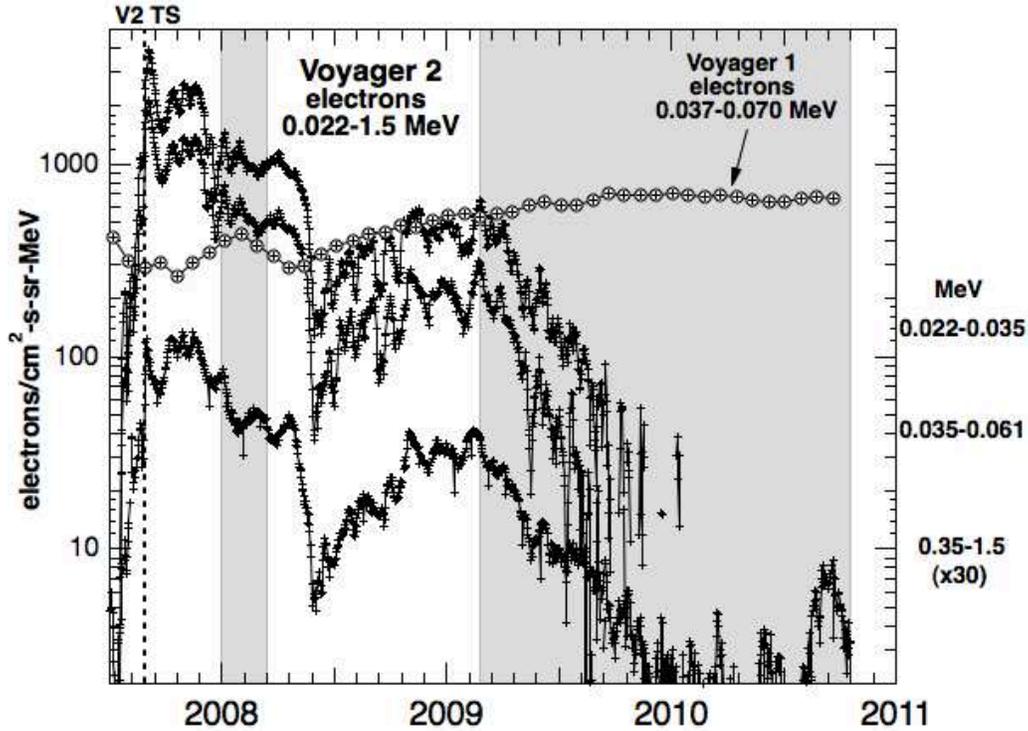}             
\caption{Daily average of the intensity of electrons from 0.022-0.035
  MeV, 0.035-0.061MeV and 0.35-1.5 MeV. In the unipolar
periods around 2008-2008.2 and from 2009.15 on (shaded areas) 
there is a drop of the intensity of electrons as measured by Voyager
2. The drop was especially dramatic after 2009.15. The
  non-shaded regions, after 2008, correspond to the sector region (the proposed
``bubble'' region). }
\label{figure5}
\end{figure}

\begin{figure}[htbp]
\centering
\includegraphics[width=0.6\textwidth]{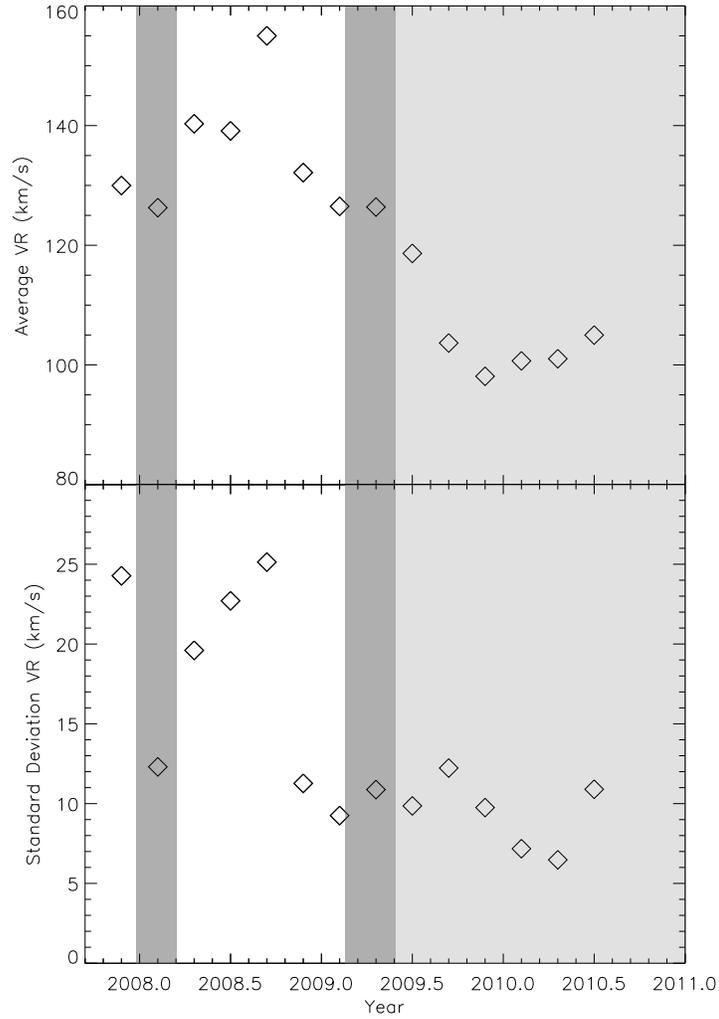}             
\caption{a) 0.2 year averages of the radial flows and b) standard deviations of the radial flows as measured by Voyager
2 from 2008.7 to 2010.5. The measured unipolar
regions (between 2008-2008.2 and 2009.15 and 2009.41) are shaded gray
and the conjectured unipolar region (after 2009.41) is shaded light
gray. The non-shaded regions
      correspond to the sector region (proposed ``bubble''
      region.)}
\label{figure6}
\end{figure}

\section{Disordered magnetic field in the Heliosheath Sector Region}

We also present results from kinetic simulations carried out with the
particle-in-cell (PIC) code p3d. The algorithm used is described in detail in Zeiler et al. 2002. Here
we present the results of simulations 
reported earlier in Drake et al. 2010. The initial state had eight
current layers, that produced a 
magnetic field $B_{x}$ with periodic reversals. Reconnection
spontaneously developed and was followed in time. 
Detailed parameters of these runs were presented in Drake et al. 2010. 

As discussed earlier in the Drake et al. 2010 paper, we argued that
the compression of the sectors on their approach to the HP would
narrow the HCS sufficiently to trigger reconnection of the sector
field. In Figures 7a and 8a we show the modulus of the magnetic
field $B$ early ($\Omega_{p}t=100$) and later ($\Omega_{p}t=150$), with
$\Omega_{p}$ the proton cyclotron frequency, to illustrate the
magnetic topology and the variability of the $B$ as reconnection
proceeds. In these simulations $y$
and $x$ correspond to the heliospheric azimuthal and radial directions
and $d_{p}$ is the proton inertial length. The ion flows in the
azimuthal direction, $v_{py}$, are shown at three times in Figure 9, the first two plots correspond to Figures 7
and 8 and the third to $\Omega_{p}t=200$. Early in time, the
driven outflows from reconnection are evident while later in time,
especially by $\Omega_{p}t=200$, the flows are erratic and do not
exhibit classical reconnection signatures.  In Figure 10 are the three
components of the ion velocity,  tangential (dotted), radial
  (solid) and normal (dashed) along a cut at $y=240d_{p}$ at a time $\Omega_{p}t=200$. This is
the same time as the 2-D plot of $v_{y}$ shown in Fig. 9c. The absence of reconnection
at late time is not because the magnetic free energy has been
completely depleted. Cuts through the simulation 
along the radial direction (at $y=240d_{p}$) in Figures 7b and 8b show
$B$ and the heliospheric elevation angle (defined as the angle between
the magnetic field and radial direction). Even late in time
substantial magnetic energy remains. Further, a casual glance at the
behavior of $\lambda$ might suggest that the sectors are still intact,
although with a more irregular spacing than earlier. These
``sectors'', however, correspond to magnetic islands, whose size is
comparable to the original sector spacing. The absence of ongoing
reconnection is a consequence of firehose stability: the interior of
these islands is at the marginal firehose condition 
($1-4{\pi}(P_{\parallel}-P_{\perp})/B^{2}=0$), where
$P_{\parallel}$ and $P_{\perp}$ represent the parallel and
perpendicular 
components of the pressure), where the magnetic tension that 
drives reconnection goes to zero (Drake et
al. 2006, 2010). Thus, the predicted late-time structure of the
sectored field consists of nested islands with irregular spacing but
with scale sizes comparable to the original sector spacing. The
remnant flows are erratic and do not exhibit the classic reconnection
signatures.

\begin{figure}[htbp]
\centering
\includegraphics[width=0.5\textwidth]{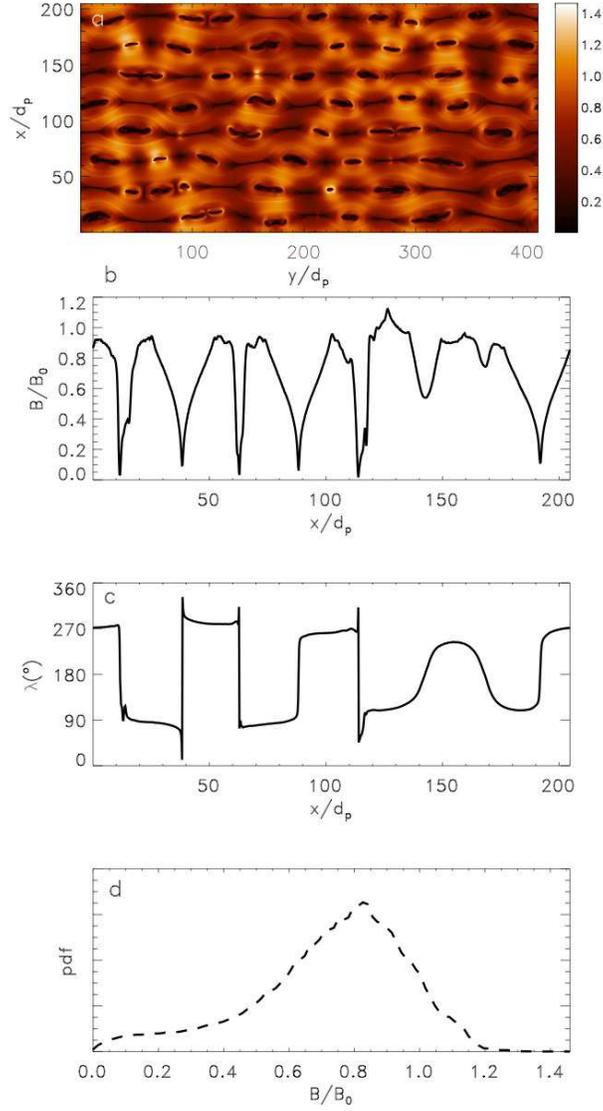}                
\caption{Magnetic field structure at $\Omega_{p}t=100$ from the PIC
  simulation described in Drake et al. (2010) where the initial state had
  $\beta=0.2$. Panels (a) show the magnitude of the magnetic field B
  during reconnection of the sectors; (b) the magnetic field intensity
  in a cut along $y=240d_{p}$; (c) angle $\lambda$ along $y=240d_{p}$
  and (d) the distribution of the magnetic field from the simulation
  (dashed line).}
\label{figure7}
\end{figure}

\begin{figure}[htbp]
\centering
\includegraphics[width=0.5\textwidth]{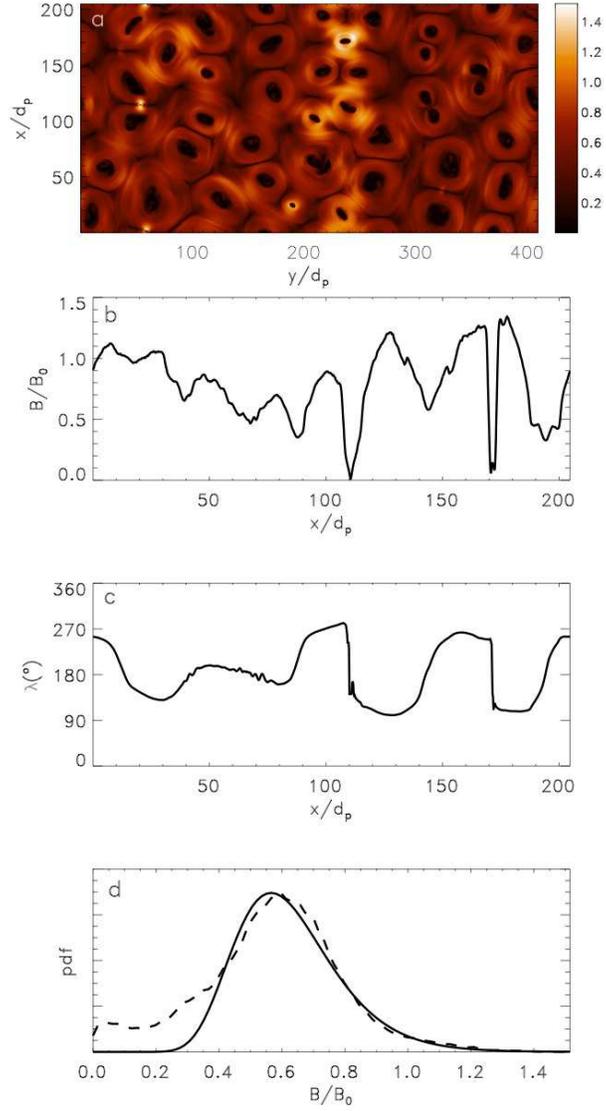}                
\caption{Same as Figure 7 but at $\Omega_{p}t=150$. The solid line in
  (d) is a log-normal distribution.}
\label{figure8}
\end{figure}

\begin{figure}[htbp]
\centering
\includegraphics[width=0.5\textwidth]{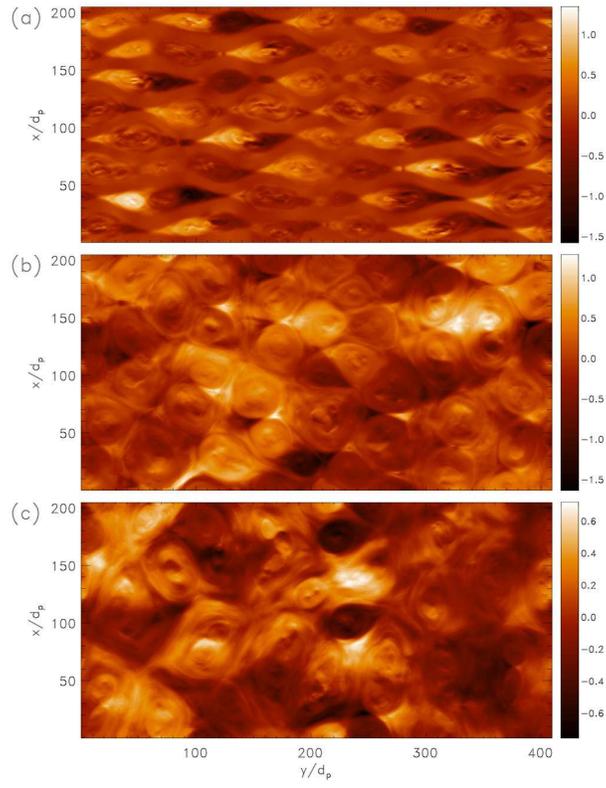}                
\caption{Flows of the protons in the azimuthal direction ($v_{py}$ at
  three times ($\Omega_{p}t=100,150,200$.)}
\label{figure9}
\end{figure}

\begin{figure}[htbp]
\centering
\includegraphics[width=0.5\textwidth]{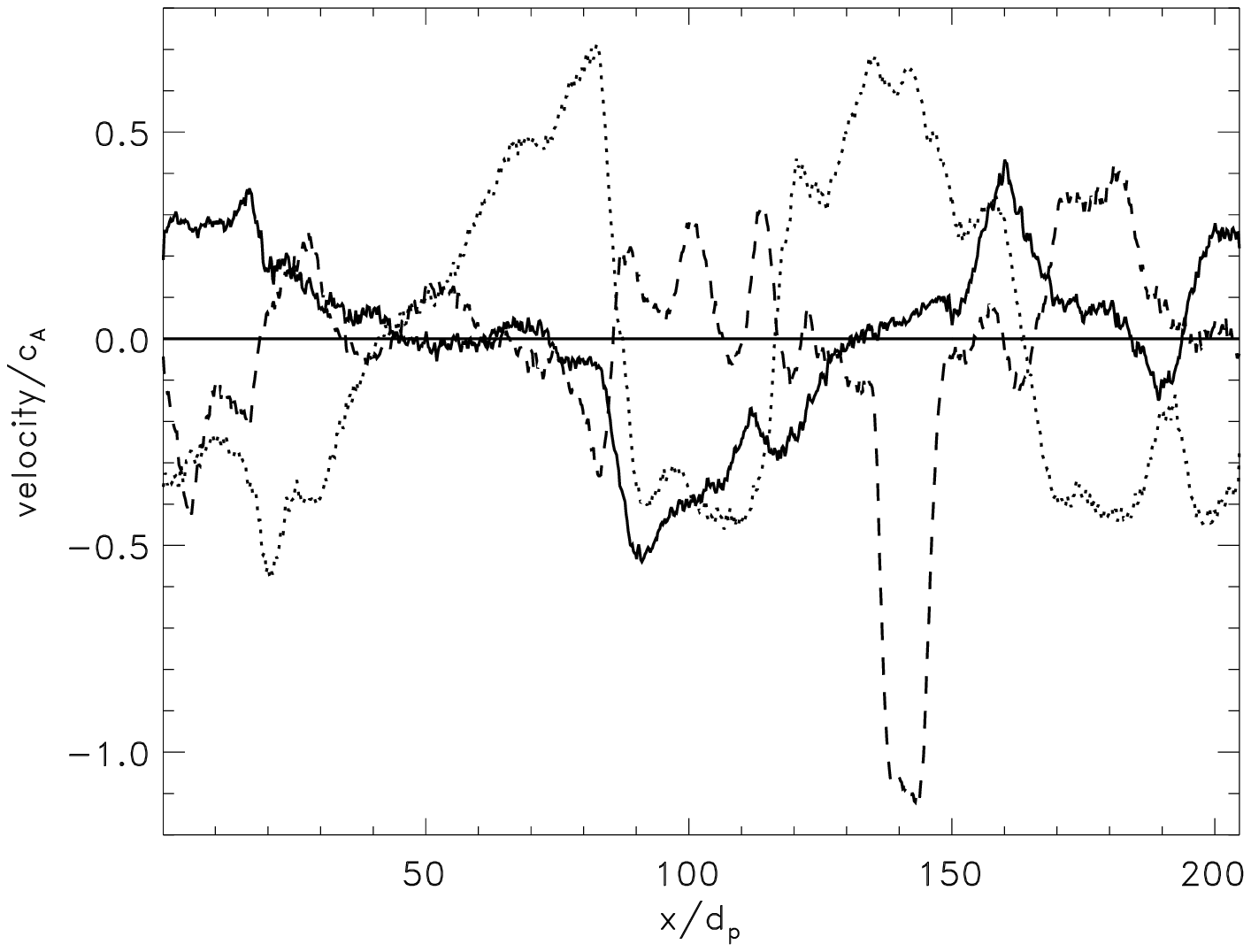}                
\caption{Three components of the ion velocity,  tangential (dotted), radial
  (solid) and normal (dashed) along a cut at $y=240d_p$ at a time
  $\Omega_{p}t=200$. 
This is the same time as the 2-D plot of $v_{y}$ shown in Fig. 9c.}
\label{figure10}
\end{figure}

A fundamental question is therefore how far upstream from the HP the
nested-island, disordered field extends. While we argued in Drake et
al. 2010 that reconnection would only onset close to the HP where the
thickness of the HCS could approach $d_{p}$, it is possible the
reconnection near the HP could spread upstream as the jostling of
reconnected islands propagates upstream.

A theoretical exploration of the upstream spread of reconnection in
the sectored field is beyond the scope of the present paper. We
argue, however, that the Voyager 2 satellite data presented in Section
III provides substantial support for the idea that the sectored field
encountered by Voyager 2 has already undergone reconnection and that
the measured magnetic field is in the disordered state shown in Figure
8.  The irregular ``sector'' structure seen in the data corresponds to
the irregular magnetic island spacing of Figure 8. The near absence
of classic reconnection signatures in the plasma flows in the Voyager
data is consistent with their absence in the simulation
data. Moreover, at least one reconnection site has been identified in
the Voyager 2 data (Burlaga \& Ness (2009)) (the presence of
{\it D-sheets} was interpreted as evidence for reconnection (Burlaga
\& Ness 1968)) that implies the existence of the ocasional reconnection
event at late time when
the magnetic islands merge. Further, the enhanced amplitude of
fluctuations in the radial flow data in the ``sectored'' region compared to
unipolar regions (Figure 6) suggests reconnection as the source of the irregular
flows. No other driver for these flows, which appear only in the
``sectored'' region, has been proposed. 

The distribution of the azimuthal angle $\lambda$ is another
  indicator of a bubble stage. We plot in Figure 11 the distribution 
of $\lambda$ for two periods: a) Days 0-65 of 2001
  where Voyager 2 was immersed in a sector region (Burlaga et
  al. 2003) upstream of the termination shock; and b) between 2008.3
  to 2009 downstream of the termination shock. We did not include data
  where the magnitude of the magnetic field was $< 0.05nT$ because, as
  noted by Burlaga et al. 2010, this is approaching the resolution limit of the magnetometer. Period (b) is representative of a period where
  we argue that the sector region is in a bubble state. We expect that, 
similar to Figure 8, $\lambda$ would be less organized than in the
regular sector region (period {\it (a)}). This can be seen in Figure
11. The distribution is broader in period {\it b} than in period {\it
  a}.

\begin{figure}[htbp]
\centering
\includegraphics[width=0.5\textwidth]{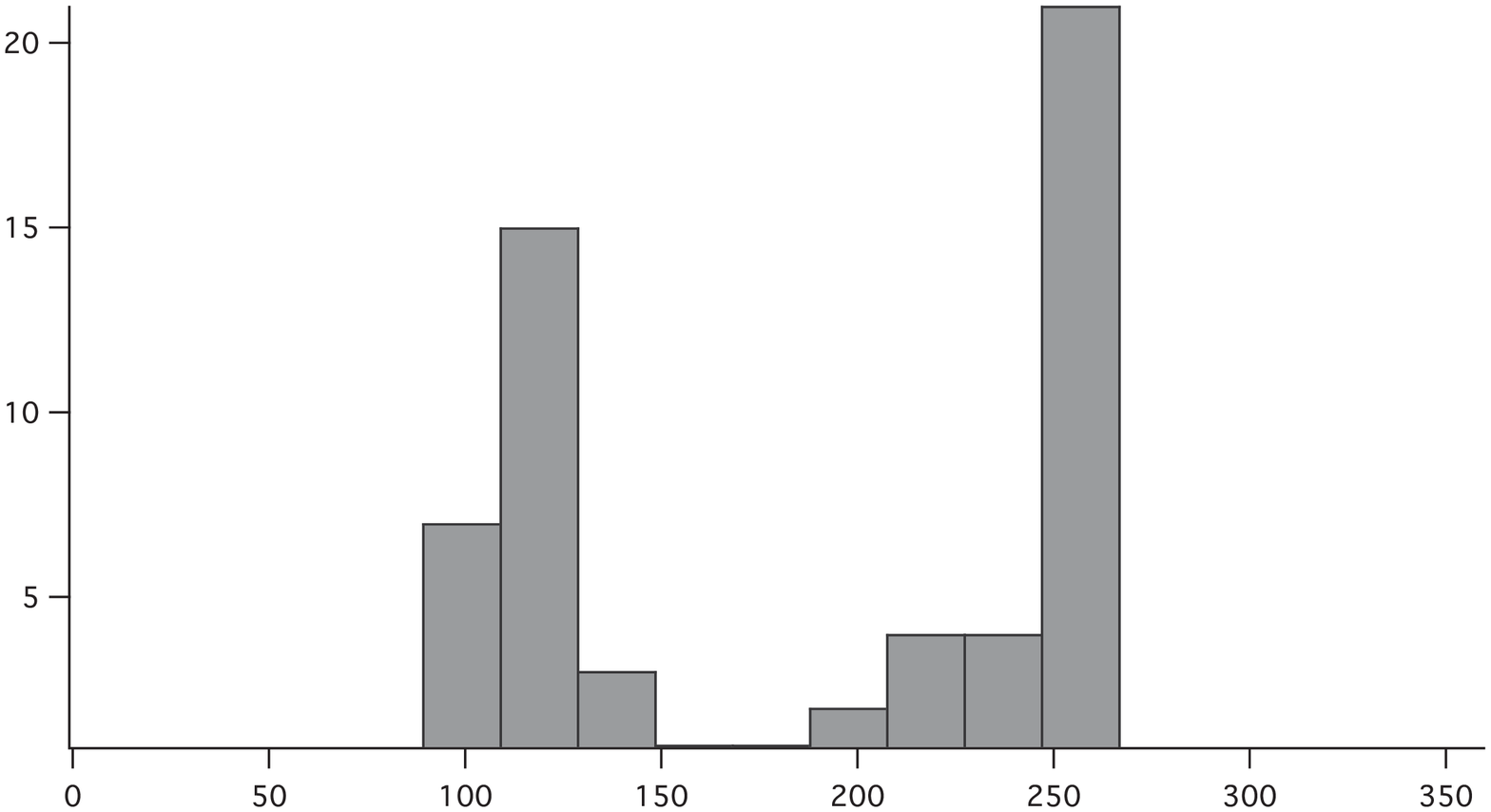} 
 \includegraphics[width=0.5\textwidth]{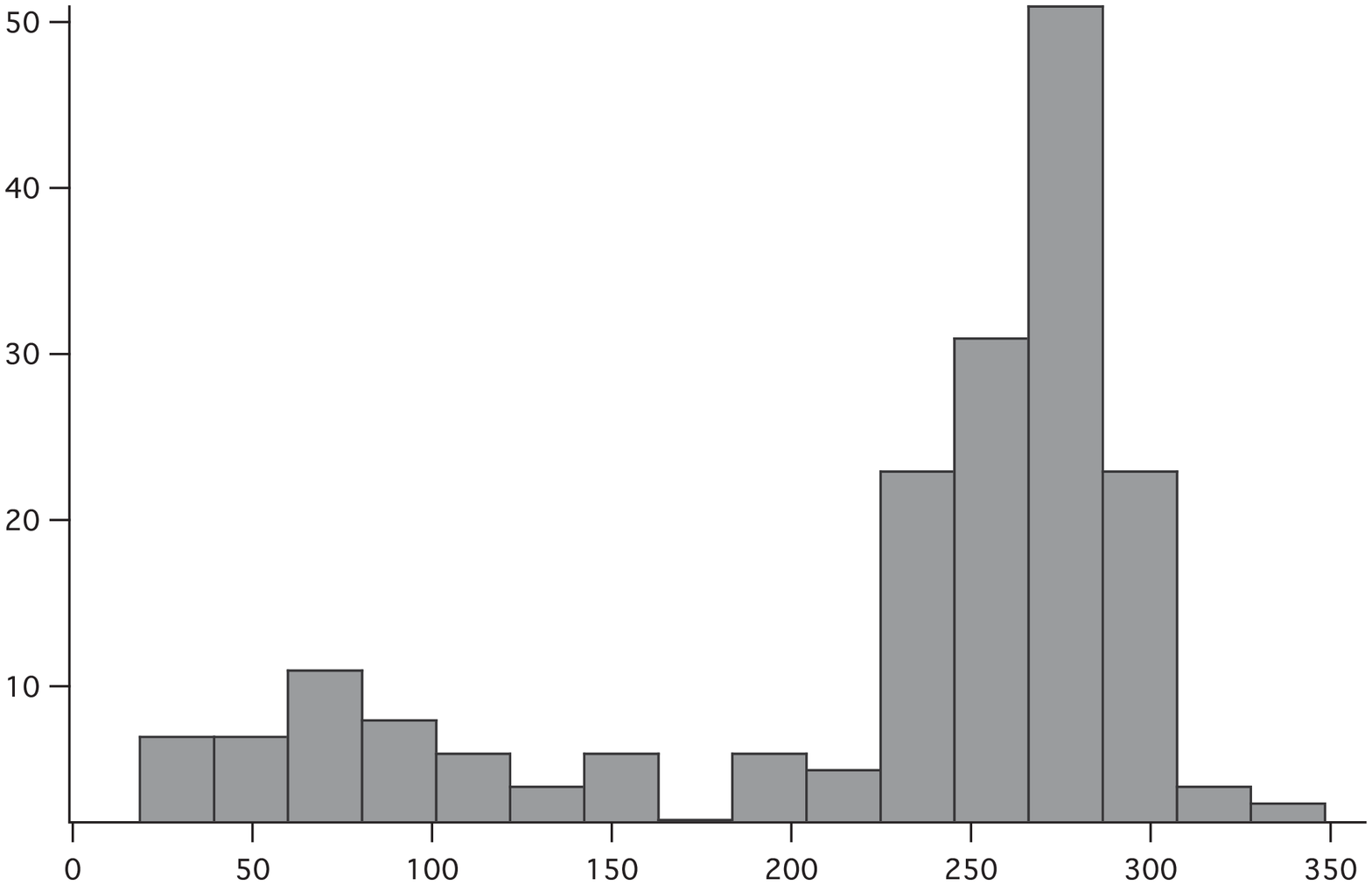}            
\caption{Distribution of the azimuthal angle $\lambda$ in a
    normal sector (a) and a bubble period (b): a) Days 0-65 of 2001
  where Voyager 2 was immersed in a sector region (Burlaga et
  al. 2003); and b) between 2008.3
  to 2009. We did not include data
  where the magnitude of the magnetic field was $< 0.05nT$ because 
this is approaching the resolution limit of the magnetometer. }
\label{figure11}
\end{figure}

A final piece of evidence in favor of a disordered magnetic field
concerns the probability distribution of the magnetic field strength
measured by Voyager 2. In the sector region the probability
distribution has been empirically found to be log-normal, while 
it appears Gaussian in
unipolar regions. The distribution of B from the PIC simulations is
shown in Figures 7d and 8d. Early in time, the distribution does not
match an obvious form. However, at late time the distribution of $B$
from the simulation (dashed line) matches a log-normal (solid line) 
surprisingly  well at large B. The match is not good at small $B$. The
discrepancy at small $B$ may be because the initial magnetic
distribution consisted of simple anti-parallel fields while in reality
the field rotates in the reversal region so there are no regions of
zero field (Smith et al. 2001). In the simulations the tail of high magnetic fields arises as the
magnetic bubbles, which exhibit substantial motion even after
reconnection has largely ended (see Figs. 9c and 10), bounce off one
another, compressing magnetic flux lying between the colliding bubbles
(white regions of Figure 8a).This behavior can be most clearly seen in
movies of the 2-D evolution of $B$ (not shown). This behavior does not take 
place during the reconnection phase of the initial current
sheets (Figures 7 b,c).

The heliosheath magnetic field data from Voyager 1 also supports the
idea that the heliosheath field in the sectored region has already
reconnected. During the 2007.4-2008.2 time period (Burlaga et al
2009b) the spacecraft was in a region of negative polarity, indicating
that it was outside of the sectored region in the region of northern
polarity. During this time the fluctuations in magnetic field strength
dropped significantly compared to time intervals when the spacecraft
was in the sectored region. The higher level of fluctuations in the
sectored region is consistent with that expected from the
nested-bubble state after reconnection of the sectored field
(Fig. 9). Finally, in 2009, in day 254 (Burlaga \& Ness (2010)) 
Voyager 1 entered into a region of
positive polarity and remained there through the remainder of
2009. Presumably the spacecraft was in the sectored region during this
period but due to the radial flow in the heliosheath remained in a
single sector. During this time period the magnetic field data
exhibited a high level of fluctuations: the magnetic field statistics
during all of 2009 exhibited a log-normal distribution (with a
pronounced tail of high magnetic field events) while during the period
1-254 exhibited a more Gaussian distribution. Again these observations
suggested that the sectored region in the vicinity of Voyager 1 is in
a disordered state.

An important question is why even when leaving 
the sector region for brief periods (e.g., during 2006.2), the intensity
of the low energy electrons did not drop at Voyager 1 (Figure 5)? We
believe that this is because the much thicker sector region in the
northern hemisphere compared with that in the southern hemisphere
(Figure 1): the thicker sector region in the north is a much stronger
source of energetic electrons than that in the south and is able to
maintain a high flux of the electrons even in the unipolar
region. Confirmation of this hypothesis will require a careful 
modeling effort taking into account the disordered field in the outer
heliosphere. 

The evolution of the reconnection in a sector field depends strongly
on the $\beta$=thermal pressure/magnetic
pressure of the initial plasma. At higher values of $\beta$
reconnection is dominated by longer wavelength modes and island
contraction quickly causes the plasma within islands to hit the
marginal firehose condition, which suppresses reconnection (Schoeffler
et al. 2011). As a result, the islands at late time in the high
$\beta$ case remain elongated rather
than nearly round as in the case of low $\beta$. This difference can
be seen by comparing the geometry of islands at late time for the
$\beta=0.2$ initial state (Fig. 8a) versus that for $\beta=4.8$
(Fig. 12a). The high $\beta$ case continues to exhibit regions of
intense magnetic field (Figs. 12a, b) and the distribution of $B$
again takes the form of a lognormal distribution (Fig. 12d). 

The round versus elogated structure of bubbles at late time has a
dramatic impact on the distribution of $\lambda$ and therefore offers
an important clue about the late time magnetic structure of the
heliosheath. A cut of $\lambda$ for $\beta=4.8$ is shown in
Fig. 12c. The differences in the $\lambda$ distributions for the low
and high $\beta$ can be most clearly seen in the distribution of
$\lambda$ averaged over the entire computational domain
(Fig. 13). Shown in Fig. 13a is the data for the $\beta=0.2$ initial
state and in Fig. 13b is that for the $\beta=4.8$ initial state. The
distribution of $\lambda$ for the $\beta=0.2$ case has peaks at
$0^{\circ}/360^{\circ}$ and $180^{\circ}$ while for the $\beta=4.8$
case it peaks around $90^{\circ}$ and $270^{\circ}$ but with a broad
distribution.  The distribution of $\lambda$ at $\beta=4.8$ is similar
to the Voyager 2 data in the hypothesized bubble region.  The value of
$\beta$ in the heliosheath is unknown, but there are indications that
it is high due to the contributions from the suprathermal population
of pickup ions.

\begin{figure}[htbp]
\centering
\includegraphics[width=0.5\textwidth]{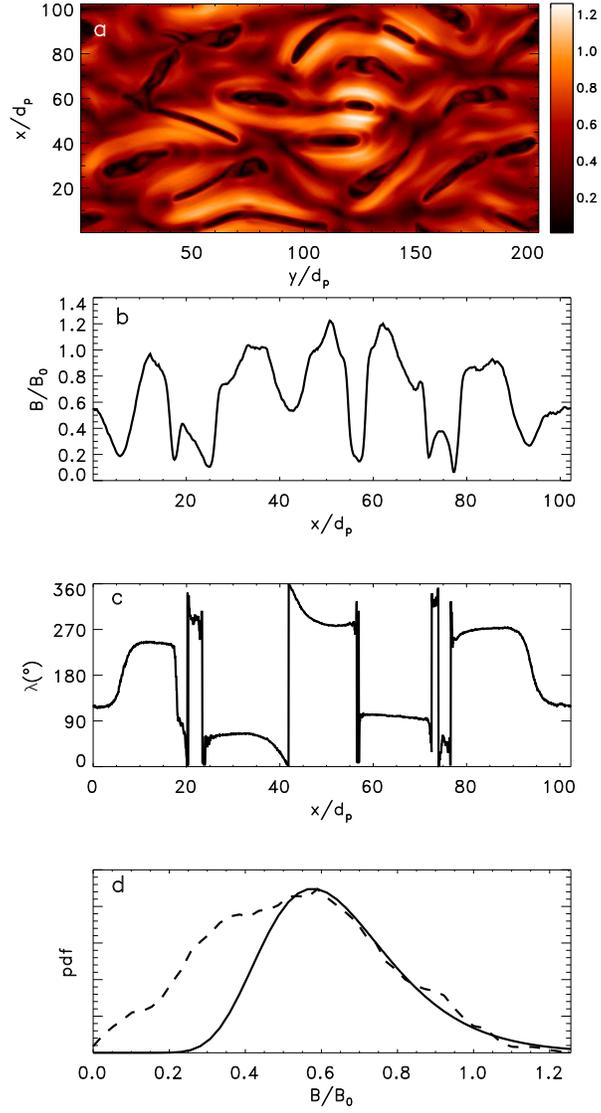}                
\caption{Same as Figure 8 for a case with $\beta=4.8$ in the initial
  state. The panels (b) and (c) show cuts along $x$ for $y=127.5d_{p}$. The solid line in (d) is a log-normal distribution.}
\label{figure12}
\end{figure}

\begin{figure}[htbp]
\centering
\includegraphics[width=0.5\textwidth]{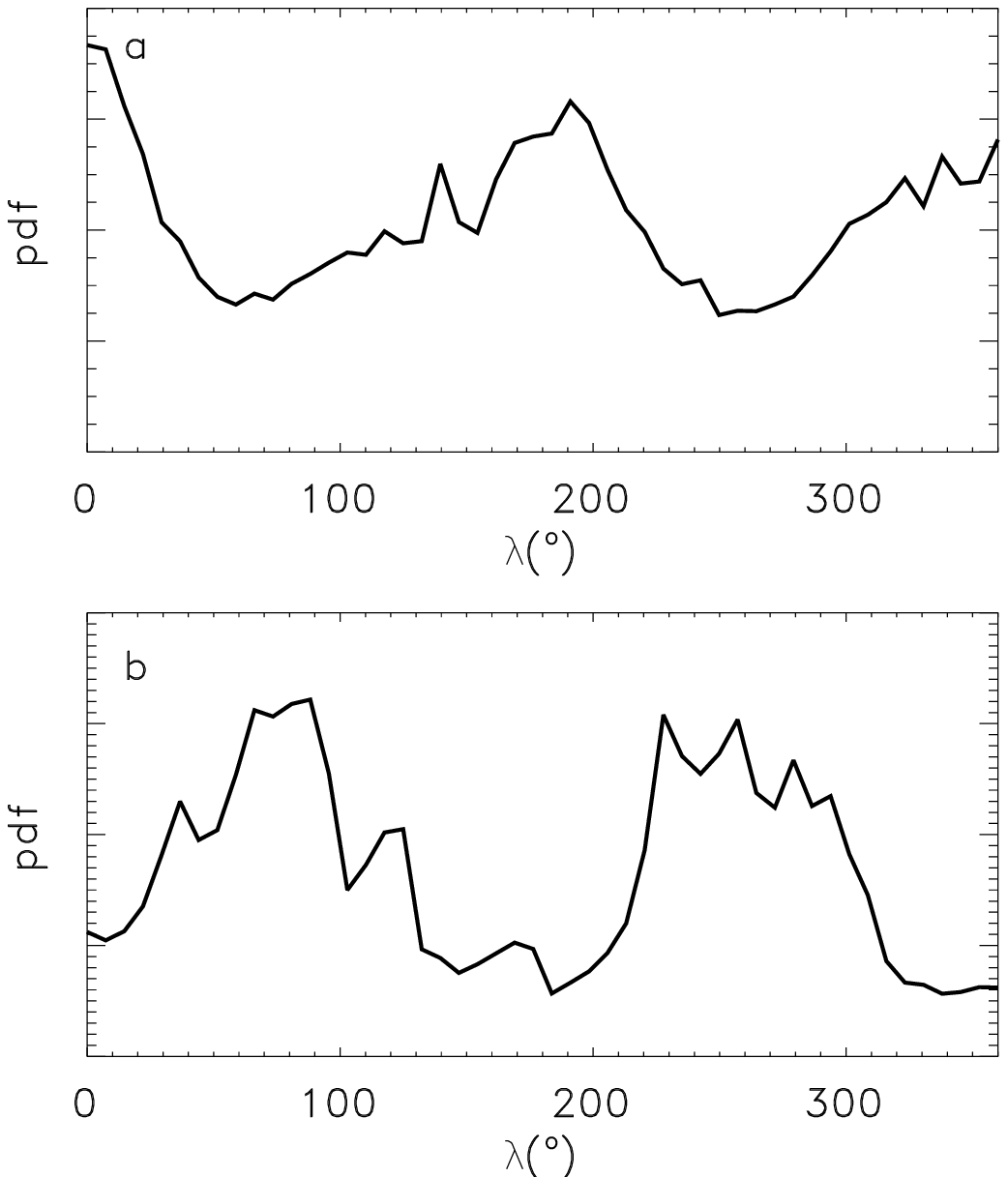}                
\caption{The distribution of  $\lambda$ over the entire simulation
  domain for initial conditions with (a) $\beta=0.2$ and (b) $\beta=4.8$.}
\label{figure13}
\end{figure}

\section{Discussion}

Our prediction is that there will be an asymmetry
between the northern and southern hemisphere with respect to the
structure of the heliospheric magnetic field close to the
heliopause. In the northern 
hemisphere and to less extent in the southern hemisphere there 
will be a disordered field resulting from reconnection of the
sectors. Additionally, we predict that
upstream of the heliopause, but within the sector region the magnetic field will
be disordered as well (see Figure 14.). The present PIC
  simulations are 2D, in the plane of the reversals of the magnetic
  field. This plane corresponds to the azimuthal plane in 3D, or in
  the xy plane in Figure 14. The structure and dynamics of the
  bubbles in 3D is an issue that will be explored in a future work.

\begin{figure}[htbp]
\centering
\includegraphics[angle=90,width=0.8\textwidth]{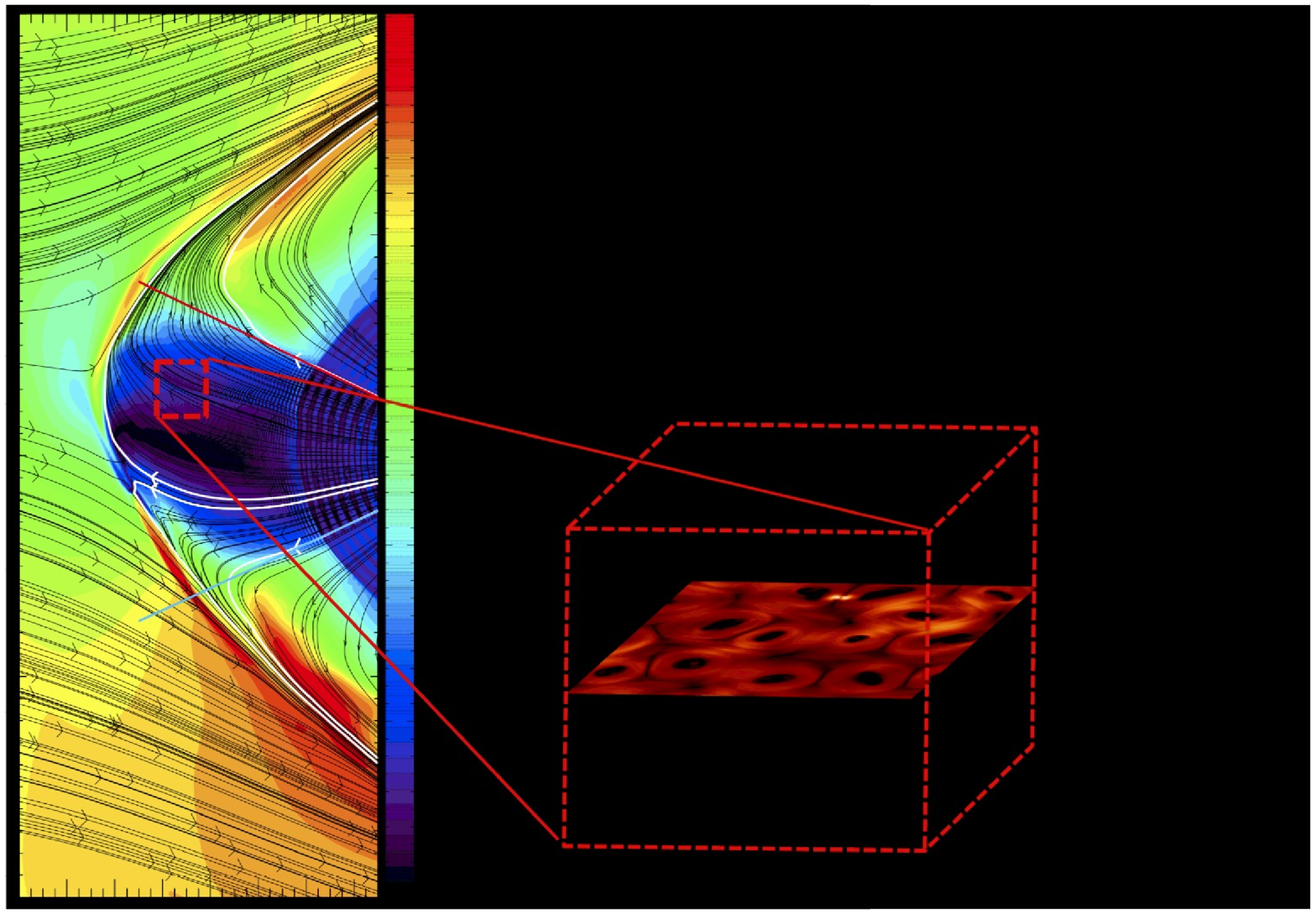}                
\caption{Representation of what we expect the outer heliosphere to be: a bath
  of bubbles. The plane from the MHD simulation is a meridional plane, a x-z plane. The PIC simulation was done in the
    azimuthal plane, the x-y plane.}
\label{figure14}
\end{figure}

Energetic electrons are especially sensitive to the topological
structure of magnetic fields.  
The presence of magnetic islands/bubbles can significantly impact the
transport of energetic electrons (and ions) in the heliosheath
compared with that in a laminar region. While in a
laminar field, the transport is dominated by field-aligned motion in
longitude; magnetic islands/bubbles in the x-y plane strongly inhibit such motion. In the
disordered field in the northern hemisphere (and the part that was
carried to the south) convective transport may compete with field aligned streaming in controlling the
spatial distribution of electrons (and ACRs). 

We predict that the electrons and ions accelerated in the heliosheath close to the heliopause will have much different transport
characteristics in a disordered sector field in comparison with the 
laminar field in the unipolar region. The electrons (and ions) will be
locally trapped on magnetic islands in a reconnected sector region and
will make their way through the heliosheath in a complex diffusion
process: magnetic islands will act as local storage vessels for
energetic particles. Once the
electrons are in a region of ordered field they will rapidly escape.
Upstream of the termination shock and at latitudes above and below
the sector boundary the field retains its nominal laminar Parker
structure. Once the electrons access the Parker field they rapidly
escape towards the inner heliosphere. The termination shock and the
sector boundaries therefore act as sinks for 
energetic electrons. We would expect the energetic electron
population in the heliosheath to be much larger than that upstream 
of the termination shock, which is what is observed. 

The observations support this scenario. From Figure 4 we argue that
Voyager 2 crossed into a region of unipolar field in the period of
2009.15 onward ($90\%$ of the time the field was negative from 2009.15-2009.4). In the period between
2008-2008.2 Voyager 2 also crossed into a unipolar region. These two
periods (shaded areas) correspond to a drop of intensity of electrons in the energy
range (0.02-1.5 MeV) measured by Voyager 2
while the
energetic electrons at Voyager 1 remained at the same level or even
increased (Figure 5).  These two periods also coincide with the periods when the
fluctuations of the radial flows as measured by Voyager 2 were low, as
seen in Figure 6. This is evidence that in these periods Voyager 2 left
a region of disordered field where they were trapped within magnetic
islands. The electrons then 
accessed the laminar Parker field and rapidly escaped. We futher
expect 
some anisotropy of electrons before and after mid 2009.15 on
the Low Energy Charged Particle Experiment (LECP).

The disordered heliospheric magnetic field near the heliopause
will affect the entrance and modulation of galactic cosmic
ray (GCR) electrons making the northern hemisphere more
``transparent''. The GCR electrons travelling along the interstellar
magnetic fields can enter and percolate through the heliosphere. The
ones entering the northern hemisphere will travel through
the disordered field of the sector region, while those in the
southern hemisphere will access a laminar field more quickly and escape. We
expect therefore a north-south asymmetry in the intensity and
modulation of the GCR electrons.

This picture is in agreement with the observations of (McDonald 2010),
which exhibit a dramatic
asymmetry between Voyager 1 and 2 in the intensity of GCR electrons
with energies of  3.8-59 MeV. The intensities at Voyager 1 continue to
rise while those at Voyager 2 reached a plateau well ``below'' that of Voyager 1. 

As described above, our scenario predicts that the transport of particles will be 
different inside and outside the sector region. In particular, it
predicts that independent 
of the solar cycle there should be an increased source of GCRs and
energetic electrons at 
lower latitudes. The energetic electrons and GCR entering the
heliosphere will percolate 
through the “bubbles” until leaving the sector region and quickly migrating into the inner heliosphere.

This scenario will affect as well our understanding of reconnection at the heliopause
where we argued that the most favorable location for reconnection was where the interstellar magnetic field was
antiparallel to the heliospheric field (Swisdak et al. 2009). In that
work we assumed that the heliosheath field was laminar with a
organized polarity in each hemisphere. What happens when a laminar interstellar
magnetic field approaches a sea of magnetic islands needs to be investigated.

The sector region is carried to the north so Voyager 1 is expected to
remain inside the sector region all the way to the
heliopause. Voyager 2 after leaving the sector region in
2009.15 is expected to eventually re-enter the sector region, but at a
radial location much closer to the heliopause.

In this work mainly we focused on the energetic electrons. However, 
we expect that because of the north-south asymmetry of 
the sector region there will be an asymmetry in the intensity of the 
ACRs between Voyager 2 in the southern hemisphere and Voyager 1 in the 
northern hemisphere. This should be more visible as the
HCS moves to lower latitudes than Voyager 2. Current analysis of the data does not reveal the 
spatial intensity gradients because the ACRs are dominated by
 temporal variations (Decker et al. 2009). Future study will have
 to disentangle the temporal variation in order to confirm our
 prediction of a gradient between Voyager 1 and 2.

\acknowledgments
M. O. would like to acknowledge the support of NASA-Voyager Guest
Investigator grant NNX07AH20G. This work is also supported by the
National Science Foundation CAREER Grant ATM-0747654.
J. D. and M. S. were supported by the grant ATM-0903964. M. O. would
like to thank especially the staff of NASA Supercomputer Division at Ames and the
Pleiades award SMD-10-1600 that allowed the simulations to be
performed. The PIC simulations were carried out at the National Energy
Research Super Computer Center (NERSC). Work at JHU/APL was supported
by the NASA Voyager Interstellar Mission, Contact NNX07AB02G. This
research 
benefited greatly from discussions that were held at the meetings of the International Team devoted to understanding the -5 tails and ACRs that has been sponsored by the International Space Science Institute in Bern, Switzerland.

\end{document}